\documentclass[11pt]{article}
\newcommand{\nn}{\noindent}
\newcommand{\no}{\nonumber\\}
\newcommand{\be}{\begin{equation}}
\newcommand{\ee}{\end{equation}}
\newcommand{\ba}{\begin{eqnarray}}
\newcommand{\ea}{\end{eqnarray}}
\newcommand{\ci}[1]{\cite{#1}}

\newcommand{\la}[1]{\label{#1}}

\def\gl#1{(\ref{#1})}

\renewcommand{\thefootnote}{\fnsymbol{footnote}}

\begin{document}

\title{\bf
 Nonlinear Supersymmetry in Quantum Mechanics: Algebraic Properties
and Differential Representation}
\author{
{\sc A. A. Andrianov}$^{a,b,}$\thanks{
andrianov@bo.infn.it}\,\, and
{\sc A. V. Sokolov}$^{b,}$\thanks{
sokolov@mph.phys.spbu.ru}\\
{$^a$  INFN, Sezione di Bologna, Via Irnerio 46,}\\
{40126 Bologna, Italy}\\
{$^b$  V.A.Fock Institute of
Physics, Sankt-Petersburg State University,}\\
 {198504 Sankt-Petersburg, Russia}}
\date{}
\maketitle

\begin{abstract}
We study the
Nonlinear (Polynomial, $N$-fold,...) Supersymmetry algebra in one-dimen\-si\-o\-nal
QM. Its structure is
determined by the type of conjugation operation (Hermitian conjugation or
transposition) and
described with the help of the Super-Hamiltonian
projection on the zero-mode subspace of a supercharge. We show that the
SUSY algebra with transposition symmetry is always polynomial in the
Super-Hamiltonian if
supercharges represent differential operators of finite order.
The appearance of the extended SUSY with several (complex or real)
supercharges
is analyzed in details and it is established that no more than two independent
supercharges
may generate a Nonlinear superalgebra
which  can be appropriately specified as ${\cal N} = 2$
SUSY. In this case we find a non-trivial hidden symmetry operator
and rephrase it
as a non-linear function of the  Super-Hamiltonian on the physical state space.
The full   ${\cal N} = 2$ Non-linear SUSY algebra includes "central charges"
both polynomial and non-polynomial (due to a symmetry operator)
in the Super-Hamiltonian.
\end{abstract}

\bigskip

\noindent
{\it PACS}:\, 03.65.Ge; 03.65.Fd, 11.30.Pb\\
{\it Key words}:\, Supersymmetric Quantum Mechanics, Nonlinear Supersymmetry,
 Quasi-solvability, Hidden symmetry.

\setcounter{footnote}{0}
\renewcommand{\thefootnote}{\arabic{footnote}}
\section{Introduction}
\hspace*{2ex}
Supersymmetric Quantum Mechanics \ci{nico,witten}
has been well proven as providing
efficient non-perturbative methods to explore
new isospectral quantum systems \ci{abi}-\ci{sukum} (see reviews
\ci{genden}-\ci{lima} and references therein) and to design nuclear
potentials with required properties \ci{baye,acd} as well as
to find  the SUSY induced hidden dynamical
symmetries \ci{ain}-\ci{new} and, more specifically,
 to search for new exactly or quasi-exactly \ci{tur,shi} solvable problems in
QM \ci{lahiri}-\ci{junker}, \ci{ain,new}, \ci{bender}-\ci{ddt}.

When being written in the fermion number representation the
one-dimensional  SUSY QM  assembles a pair of
isospectral Hamiltonians
$h^+$ and $h^-$ into the matrix Schr\"odinger operator,
a Super-Hamiltonian,
\ba
H = \left(\begin{array}{cc}
h^+& 0\\
0 & h^-
\end{array}\right) =\left(\begin{array}{cc}
-\partial^2 + V_1(x)& 0\\
0 & - \partial^2 + V_2(x)
\end{array}\right) \equiv -\partial^2 {\bf I} + {\bf V}(x) , \la{suham}
\ea
where $\partial \equiv d/dx$.
The isospectral connection between
components of the Super-Hamiltonian is provided by  intertwining
relations with the help of Crum-Darboux
(see \ci{matveev} and references therein) differential  operators  $q^{\pm}$,
\be
h^+ q^+ = q^+ h^- , \quad q^- h^+ = h^- q^- , \la{intertw}
\ee
which, in the framework of SUSY QM,  are components of the supercharges,
\ba
Q=\left(\begin{array}{cc}
 0 &  q^+\\
0 & 0
\end{array}\right),\quad
\bar Q=\left(\begin{array}{cc}
 0 &  0\\
q^- & 0
\end{array}\right),\quad  Q^2 = \bar Q^2 = 0. \la{such}
\ea
The isospectral shift \gl{intertw} entails the
conservation of supercharges or the supersymmetry of the Super-Hamiltonian,
\be
[H,Q] = [H,\bar Q] =0, \la{cons}
\ee
which represents the basis of the SUSY algebra.
However its algebraic closure
is given, in general, by  a non-linear SUSY relation,
\be
\left\{Q,\bar Q\right\} = {\cal P} (H), \la{poly}
\ee
where ${\cal P} (H)$ is a function of the Super-Hamiltonian.
In this extended form the Polynomial (or
Higher-derivative)
 SUSY algebra was systematically
introduced\footnote{In a different context certain
higher derivative SUSY charges can be also associated with higher-order
Darboux-Crum transformations \ci{matveev}.} in \ci{ais,acdi}:
its supercharges were realized by certain
$N$-th order differential operators,
\be
q^{\pm}_N =\sum_{k=0}^N w^{\pm}_k (x)\partial^k, \quad w^{\pm}_N = const
\equiv (\mp 1)^N.
\la{superch}
\ee
The coefficient functions $w^{\pm}_k (x)$ are,
in general,  complex and sufficiently smooth. If they are real
then from the
hermiticity of the Hamiltonians and from Eqs.~\gl{intertw} it follows that,
$q^- \equiv (q^+)^\dagger = (q^+)^t$ with the notations:
$^\dagger$ for
the Hermitian conjugation and $^t$ for transposition.
But in the complex case the four types of SUSY algebra can be introduced,
based on four intertwining operators and, respectively, four supercharges,
\ba
&&q^- = (q^+)^\dagger,\quad \bar Q = Q^\dagger,\no
&&q^-_c = (q^+)^t,\quad \bar Q_c = Q^t,\no
&&q^+_c = (q^-)^t = (q^+)^*,\quad  Q_c = Q^*, \la{htsym}
\ea
where $^*$ obviously stands for the complex conjugation of
coefficient functions.
These algebras are generated by the pairs:
\ba
&&{\cal A}_1 = (\bar Q, Q),\qquad
{\cal A}_ 2 =(\bar Q_c, Q),\no
&&{\cal A}_3 =(\bar Q, Q_c),\qquad{\cal A}_4 = (\bar Q_c, Q_c).\la{4susy}
\ea
These sets are
eventually united in the complex, nonlinear ${\cal N} = 2$
supersymmetry (see Sec.~5).

Recently the Polynomial (or Higher-derivative)
SUSY algebra has attracted much interest \ci{bags}-\ci{fermun}
 being a natural algebraic realization of the ladder \ci{abi,sukum}
or dressing chain \ci{shab,adler} algorithms.
It was rediscovered
under the name of Nonlinear SUSY \ci{klip2,ply,klip} and of $N$-fold SUSY
\ci{akosw,asty,ast,ast2,anst,tan}.
Perhaps the label of Nonlinear SUSY (which we adopt further on)
serves better to reflect its essence
as one
could easily produce non-polynomial examples by means of limiting procedure
applied to a Polynomial SUSY algebra when supercharges would become
pseudo-differential operators or by considering SUSY algebras
with complex-valued supercharges (Sec.~3 and 5).

Meantime it was claimed in \ci{ast2,tan} that
there exists a $N$-fold SUSY which generalizes the Polynomial SUSY
in a non-trivial way. To be precise the theorem formulated and proven in
\ci{ast2} states the following.\\

\noindent
\underline{Aoyama-Sato-Tanaka (AST) theorem}\\
{\it
Let $\phi^{\pm}_{n}(x)$ $(n=1,\cdots N)$
be two sets of $N$ linearly independent functions,  zero-modes of the
supercharge
components \gl{superch},
\begin{eqnarray}
q^{\pm}_N \phi^{\pm}_{n}=0,\quad q^-_N =(q^+_N)^\dagger. \la{zerom}
\end{eqnarray}
Then the following propositions hold\footnote{\rm The first proposition is a
necessary condition for the Hamiltonian system to be quasi-exactly solvable
 \ci{tur,shi} and it was investigated recently
\ci{zhdan,fush,doeb} within the notion of ``conditional symmetry''. }

\noindent
1) The Hamiltonians $h^{\pm}$ have  finite matrix representations
when acting on the set of functions $\phi^{\pm}_{n}(x)$,
\begin{eqnarray}
h^{\pm}\phi^{\mp}_{n}=\sum_{m} \bar S^{\pm}_{nm}\phi^{\mp}_{m}.
\la{quasih}
\end{eqnarray}

\noindent
2) With the help of the $N \times N$
matrices $\bar{\bf S}^{\pm}$, the SUSY algebra closure\footnote{The
``Mother Hamiltonian'' in the terminology of
\ci{ast,ast2}} takes the general form\footnote{It is just an algebra of ${\cal A}_1$ type with
$\bar Q = Q^\dagger$.},
\begin{eqnarray}
\left\{Q, Q^\dagger\right\} =
\left(
\begin{array}{cc}
{\rm det}{\bf M}^{+}_N(h^{+})+ \pi^+_{1}q_N^-  & 0\\
0 & {\rm det}{\bf M}^{-}_N(h^{-})+ \pi^{-}_{2}q^+_N
\end{array}
\right),
\la{ASTrep}
\end{eqnarray}
where
\ba
Q =\left(\begin{array}{cc}
 0 &  q^+_N\\
0 & 0
\end{array}\right) , \quad
{\bf M}^{\pm}_N (E)\equiv E{\bf I}- \bar{\bf S}^{\pm},
\la{matrM}
\ea
and $\pi^{\pm}_{1,2}$ are differential operators of  lower orders
$N_{1,2} \leq N-1$
generating at least one more SUSY algebra (intertwining relations) for
the {\it same} Super-Hamiltonian $H$, Eq.~\gl{suham},
$[H, P_{1,2}] = [H,  \bar P_{1,2}] =0$
with supercharges,
\begin{eqnarray}
P_{1,2} =
\left(
\begin{array}{cc}
0 & \pi^+_{1,2}\\
0 & 0
\end{array}\right);\quad \bar P_{1,2} =\left(P_{1,2}\right)^\dagger;\quad
\pi^-_{1,2} = (\pi^+_{1,2})^\dagger .\la{pchar}
\end{eqnarray}

\noindent
3) In the case of non-vanishing $\pi^\pm_{1,2}$  the SUSY algebra \gl{ASTrep}
coexists at least with one Polynomial SUSY of lower order $N_* \leq
N_{1,2} < N$.

\noindent
4) If for given
$h^{\pm}$ the $N$-fold supercharges are {\it uniquely} determined then
  $\pi^{\pm}_{1,2}$ must be zero and the superalgebra closure
leads to the Polynomial SUSY of order $N$,
 \begin{eqnarray}
\left\{Q, Q^\dagger\right\} = {\rm det}{\bf M}_N^{+}(H)
={\rm det}{\bf M}_N^{-}(H) \equiv {\cal P}_N (H). \la{polsusy}
\end{eqnarray}
}

\medskip

Certainly the AST theorem is helpful to make links to quasi-exact solvability
of particular quantum Hamiltonians \ci{sasaki}. However it
does not explain the origin of generators
of ``small'' supersymmetries $P_{1,2}$.
It does not elucidate also the relationship
between them and between the matrices
$\bar{\bf S}^+$ and $\bar{\bf S}^-$.
As well it does not give a hint on what is the
maximal order of a coexisting Polynomial SUSY and how many supercharges may
commute with a given Super-Hamiltonian. From \gl{ASTrep}
it is evident that the operators  $\pi^+_1 q_N^-$ and
$\pi^-_2 q_N^+$ are related to
conserved symmetries of the Hamiltonians $h^\pm$. However neither genuine
hermitian symmetry generators nor the very meaning of such symmetries have
not been obtained in the framework of the AST theorem.

The main aim of the present work is to clarify the above mentioned,
missing points and
furthermore to prove the following.
\nn
\begin{enumerate}
\item For a given pair of isospectral systems
intertwined by differential operators of order $N$
there is always a choice of certain
intertwining operators with real coefficients
(not necessarily unique) which lead to
supercharges of a Polynomial SUSY of the same order $N$.
Thereby the $N$-fold SUSY of \ci{ast,ast2}
 for a given
quantum system {\it always} coexists with a Polynomial SUSY of the same order
(and possibly few other polynomial SUSY of different orders).
\item The complex extension of Nonlinear  SUSY
may bring a SUSY algebra different from a polynomial one just containing
non-trivial symmetry operators. These differential operators of odd order  can
be replaced by non-polynomial functions of a Super-Hamiltonian being defined in the
Hilbert space spanned on eigenfunctions of the Super-Hamiltonian. Thus for the same
Super-Hamiltonian one can simultaneously introduce the nonlinear SUSY in both a
polynomial and a non-polynomial form. In particular it
covers the propositions of the AST theorem.
\item Among the (infinite) variety of supercharges of type $Q$ (or of type 
$\bar Q$) commuting with a given
Super-Hamiltonian one can systematically find the optimal set of (no more than)
two basic SUSY
generators  
which are differential operators of even and odd order with
real coefficients. Respectively
all other supercharges of type $Q$ (or of type 
$\bar Q$) represent linear combinations of  basic
supercharge(s) with coefficient(s) polynomial in the Super-Hamiltonian.
For two essentially independent supercharges a dynamical  symmetry 
for the Super-Hamiltonian arises and the related symmetry 
operator is
unique up to a multiplier polynomial in the Super-Hamiltonian.
\item There is
a more efficient formulation of the AST theorem
which manifestly  uses the emerging
 dynamical symmetry for a Super-Hamiltonian with two supercharges
and uniquely specifies the relationship between $q^\pm$,
$\pi^\pm_1$ and $\pi^\pm_2$ and between the matrices
$\bar{\bf S}^+$ and $\bar{\bf S}^-$.
\item For isospectral systems with two independent
supercharges the notion of irreducibility for Polynomial SUSY
formulated in \ci{acdi} does not characterize firmly potentials and
the same system may be well described by a more reducible and less reducible
SUSY algebra.
\end{enumerate}

All theorems and constructions are exemplified by means of an exactly solvable
system of second order.

\section{Superalgebras with transposition symmetry}
\hspace*{2ex} The superalgebras with real coefficient functions in the
differential representation of supercharges as well as the ${\cal A}_{2,3}$
complex superalgebras have the transposition symmetry, $\bar Q_c = Q^t$.
The following theorem is valid for these superalgebras (compare with the
AST theorem).\\

\noindent
\underline{Theorem on SUSY algebras with T-symmetry}\\
{\it
Let us again introduce two sets of $N$ linearly independent functions\\
$\phi^{\pm}_{n}(x)$
$(n=1,\cdots N)$  which represent complete sets of zero-modes
of the supercharge components \gl{superch},
\begin{eqnarray}
q^{\pm}_N \phi^{\pm}_{n}=0,\quad q^-_N =(q^+_N)^t. \la{tmode}
\end{eqnarray}
Then:\\
\noindent
1) the Hamiltonians $h^{\pm}$ have finite matrix representations
when acting on the set of functions $\phi^{\pm}_{n}(x)$,
\begin{eqnarray}
h^{\pm}\phi^{\mp}_{n}=\sum_{m} S^{\pm}_{nm}\phi^{\mp}_{m},
\la{quasiham1}
\end{eqnarray}

\noindent
2) the SUSY algebra closure with $\bar Q_c = Q^t$
takes the polynomial form,
\begin{eqnarray}
\left\{Q, Q^t\right\} =
{\rm det} \left[E{\bf I}-{\bf S}^{+}\right]_{E = H}
={\rm det} \left[E{\bf I}-{\bf S}^{-}\right]_{E = H} \equiv
{\cal P}_N (H),\la{tposusy}
\end{eqnarray}
irrespectively on whether the supercharge of order $N$ is unique or there exist
several supercharges for a given Super-Hamiltonian $H$.}

\bigskip

We stress that the matrix ${\bf S}^{-}$ is the same as in the AST theorem,
${\bf S}^{-}= \bar{\bf S}^{-}$
whereas the matrix ${\bf S}^{+}$ is different\footnote{From
the definitions \gl{zerom} and \gl{tmode} it follows that the spectra of
two matrices ${\bf S}^{+}$ and $\bar{\bf S}^{+}$
are mutually complex conjugated.} from  $\bar{\bf S}^{+}$
due to Eq.~\gl{tmode}.

The {\it proof} of the first statement of the theorem is analogous to that one
of the AST theorem. Namely, one has to act by the operator intertwining
relations \gl{intertw} on the zero-mode functions $\phi^{\pm}_{n}(x)$,
\begin{eqnarray}
q^{\pm}_N  h^\mp \phi^{\pm}_{n} = h^\pm q^{\pm}_N \phi^{\pm}_{n}=0,
\end{eqnarray}
and conclude therefrom that $h^\mp \phi^{\pm}_{n}$ is also a
zero-mode solution, i.e. can be expressed as a linear
combination \gl{quasiham1}  of a complete set of
$\phi^{\pm}_{n}(x)$.

The {\it proof} of the second part of this theorem is based
on the properties assembled into the
lemma.\\

\noindent
\underline{Lemma}\\
{\it
Let $\lambda^\pm_1,...,\lambda^\pm_N$ be two sets of eigenvalues of matrices
${\bf S}^{\pm}$ as being introduced in the above formulated theorem.
Then there exist two sets of first-order differential
operators $r^\pm_1,...,r^\pm_N$  such that:\\
1) they have the canonical form,
\be
r^\pm_l = \mp \partial + \chi^{\pm}_l(x),
\quad l = 1,...,N, \la{factop}
\ee
where the functions  $\chi^{\pm}_l(x)$ may
be complex and/or singular at some points;

\noindent
2) the factorizations hold,
\be
q^+_N = r^+_1 \cdots r^+_N,\quad
q^-_N = r^-_1 \cdots r^-_N; \la{fact}
\ee

\nn
3) the chain relations take place,
\ba
(r^\mp_l)^t\cdot r^\mp_l  + \lambda^\pm_l& =&  r^\mp_{l+1}\cdot
(r^\mp_{l+1})^t +
\lambda^\pm_{l+1} \equiv h^\pm_l,\quad l = 1,\ldots,N-1\no
 (r^\mp_{N})^t \cdot r^\mp_{N} + \lambda^\pm_{N}& =& h^\pm \equiv h^\pm_N,\no
r^\mp_1 \cdot (r^\mp_1)^t  + \lambda^\pm_1& =&  h^\mp \equiv h^\pm_0,
\la{chain}
\ea

\nn
4) the intermediate Hamiltonian operators have the Schr\"odinger
form,
\be
 h^\pm_l = - \partial^2 + v^\pm_l (x); \quad  v^\pm_l (x) =
(\chi^\mp_l(x))^2 \mp
(\chi^\mp_l(x))' + \lambda^\pm_l=
(\chi^\mp_{l+1}(x))^2 \pm
(\chi^\mp_{l+1}(x))' + \lambda^\pm_{l+1},
\ee
but, in general, with complex and/or singular potentials;

\nn
5) the intertwining relations are valid,
\be
h^\pm_{l-1} \cdot r^\mp_l =  r^\mp_l\cdot h^\pm_{l},\quad  (r^\mp_l)^t\cdot h^\pm_{l-1} =
h^\pm_{l} \cdot
(r^\mp_l)^t. \la{chaint}
\ee
}

The {\it proof} of this lemma is based on the quasi-diagonalization
of matrices ${\bf S}^{\pm}$,
i.e on their reduction to the Jordan canonical form  $\widetilde{\bf S}^{\pm}$
which is block-diagonal and contains
the Jordan cells with eigenvalues on the main diagonal and unities on the upper subdiagonal.
This diagonalization can be realized by nondegenerate linear
transformations $\Omega^\pm$ of the zero-mode sets
which induce the similarity transformations of matrices ${\bf S}^{\pm}$,
\ba
\tilde\phi^\pm_l &=& \sum_{m=1}^N \Omega^\mp_{lm} \phi^\pm_m,\quad
h^\pm \tilde\phi^\mp_l =  \sum_{m=1}^N \widetilde{S}^{\pm}_{lm}
\tilde\phi^\mp_m,\no
\widetilde{\bf S}^{\pm}
&=&  \Omega^\pm {\bf S}^\pm \left(\Omega^\pm\right)^{-1}. \la{jord}
\ea
Certainly it is  sufficient to
elaborate the factorization of the operator $q^-_N$.
The last line in the matrix $\widetilde{\bf S}^+$ contains the Hamiltonian
eigenvalue $\lambda^+_N$,
\be
h^+ \tilde\phi^-_N = \lambda^+_N \tilde\phi^-_N. \la{hameig}
\ee
Next we define,
\be
r^-_N \tilde\phi^-_N = 0, \quad \chi^-_N \equiv
- \frac{\left(\tilde\phi^-_N\right)'}{\tilde\phi^-_N}. \la{mode1}
\ee
From Eq.~\gl{hameig} it follows that
\be
h^+ \equiv h^+_N  = (r^-_N)^t\cdot r^-_N + \lambda^+_N.
\ee
Furthermore, the intermediate Schr\"odinger-like Hamiltonian can be introduced,
\be
h^+_{N-1} =  r^-_N \cdot (r^-_N)^t + \lambda^+_N = - \partial^2 + (\chi^-_N)^2 +
(\chi^-_N)' +
\lambda^+_N,
\ee
which is obviously involved in the intertwining relations with $h^+_N$,
\be
h^+_{N-1} \cdot r^-_N =  r^-_N \cdot h^+_{N},\quad  (r^-_N)^t \cdot h^+_{N-1} =
h^+_{N}\cdot
(r^-_N)^t.
\ee

The combination of Eq.~\gl{tmode} and Eq.~\gl{mode1} yields the
factorization,
\be
q^-_N = q^-_{N-1} \cdot r^-_N.
\ee
Indeed if \gl{tmode} and \gl{mode1} are valid then in \gl{superch}
\be
 w^{-}_0 (x) = \sum_{k=1}^{N} w^{-}_k (x)\, \xi_k (x),\quad
\xi_k (x) \equiv (\hat\partial-
\chi^-_N)^{k-1}
\chi^-_N,
\la{wminus}
\ee
where we have introduced the notation $\hat\partial$
for the derivative of coefficient
functions  to make a clear distinction from the differential
operator $\partial$.
Therefore
\be
q^{-}_N =\sum_{k=1}^N w^{-}_k (x)\left(\partial^k +
(\hat\partial - \chi^-_N)^{k-1} \chi^-_N\right).
\ee
Respectively the factorization
is realized in each component,
\be
\left(\partial^k +
(\hat\partial - \chi^-_N)^{k-1} \chi^-_N\right) = \left[\partial^{k-1}  -
\sum_{m=0}^{k-2}\partial^{k-m-2}  \xi_{m+1}  \right](\partial +
\chi^-_N).
\ee
Now it is straightforward to show that:\\
1) the intertwining relation holds,
\be
h^- \cdot q^-_{N-1} = q^-_{N-1}\cdot h^+_{N-1},
\ee
2) the functions,
\be
\psi_n^- = r^-_N \tilde\phi^-_n, \quad 1 \leq n \leq N - 1,
\ee
form the complete, linearly independent set of solutions of the equation
$q^-_{N-1}\psi =0 $;\\
3) the matrix $\widetilde{\bf S}^+_{N-1}$ which is uniquely
determined from relations,
\be
 h^+_{N-1} \psi_n^- = \sum^{N-1}_{m=1}
\left(\widetilde{S}^+_{N-1}\right)_{nm} \psi_m^-,
\ee
in fact, is derived from $\widetilde{\bf S}^+$
after deleting of the last column and the
last line: thereby the matrix
$\widetilde{\bf S}^+_{N-1}$ still has
the Jordan canonical form and its spectrum consists of
$\lambda^+_1,\ldots,\lambda^+_{N-1}$.

The first two statements are direct consequences of the basic relations
\gl{intertw} and \gl{tmode} respectively, whereas the third one can be
obtained when acting by the operator $r_N^-$ on the definition
\gl{jord} of the matrix $\widetilde{\bf S}^+$.

Thus we have reduced the factorization problem of order $N$ to the
latter one of order $N-1$ having
proved the statements of the lemma on this step. Evidently one can proceed
recursively further on and prove completely the lemma by induction.\\

\bigskip

In turn, the {\it proof} of the above formulated Theorem uses
the factorization and intertwining relations provided by the Lemma,
\ba
q^+_N \cdot q^-_N &=& (r^-_N)^t \cdots (r^-_1)^t\cdot r^-_1 \cdots r^-_N =
(r^-_N)^t
\cdots (r^-_{2})^t (h^+_{1} - \lambda^+_1) r^-_{2} \cdots r^-_N\no
&=&  (r^-_N)^t \cdots (r^-_{2})^t \cdot r^-_{2}  \cdots r^-_N
(h^+  - \lambda^+_1)
= \cdots =  (h^+ - \lambda^+_1)\cdots (h^+ - \lambda^+_N)
\no
&=& \mbox{\rm det} \left[E{\bf I}-{\bf S}^{+}\right]_{E = h^+};\no
q^-_N \cdot q^+_N &=& r^-_1 \cdots r^-_N \cdot (r^-_N)^t \cdots (r^-_1)^t =
r^-_1 \cdots r^-_{N-1} (h^+_{N-1} - \lambda^+_N) (r^-_{N-1})^t \cdots
(r^-_1)^t\nonumber\\
&=& \cdots =  (h^- - \lambda^+_1)\cdots (h^- - \lambda^+_N)
= {\rm det} \left[E{\bf I}-{\bf S}^{+}\right]_{E = h^-}.
\la{trsusy}
\ea
The same relation can be derived for the factorization of the operator
$q^+_N$ into a product of $r^+_l$ using the Jordan form for the matrix
$\widetilde{\bf S}^-$. It leads to the equivalent representation
of the polynomial algebra,
\be
q^\pm_N\cdot q^\mp_N = (h^\pm - \lambda^-_1)\cdots (h^\pm -
\lambda^-_N)= {\rm det} \left[E{\bf I}-{\bf S}^{-}\right]_{E = h^\pm}.
\la{trsusy1}
\ee
As the relations \gl{trsusy} and \gl{trsusy1} are operator ones they hold
for any values of
spectral parameter $h^\pm \psi = \epsilon \psi$. Therefore the
eigenvalues of matrices ${\bf S}^+$ and ${\bf S}^-$ and their corresponding
degeneracies
coincide.\\

\bigskip

In general, the eigenvalues  of matrices ${\bf S}^\pm$ are complex. For real components of supercharges $q^\pm_N$, the complex
eigenvalues obviously appear in complex conjugated pairs providing the real polynomial  ${\cal P}_N (x)$ but for complex
supercharge components $q^\pm_N$ the resulting polynomial contains
complex coefficients. In next sections we examine the non-uniqueness
of a complex supercharge describing a given hermitian Super-Hamiltonian.

\section{Several supercharges and Extended SUSY}
\hspace*{2ex}  The non-uniqueness of supercharges
was  mentioned for the first time\footnote{ See the related
Section in the E-archive version of \ci{comalg} as it was eliminated from
the final journal paper under the severe pressure of a referee.} in
\ci{comalg}.
It was observed that for a hermitian Super-Hamiltonian $H$ the conserved
supercharges $Q, \bar Q$ with complex intertwining
components $q^\pm_N$ always generate
two SUSY algebras: one for their ``real'' parts $K, \bar K $  and another one
for their ``imaginary'' parts $P, \bar P$
where the corresponding labels are referred
to the real and imaginary parts of
coefficients
in the differential intertwining operators
$q^\pm_N = k^\pm_{N} + i p^\pm_{N_1}$,
\be
Q = K + i P ;  \quad [H, Q] =  [H, K] = [H, P] =
[H, \bar Q] = [H, \bar K] = [H, \bar P]
= 0. \la{extsusy}
\ee
Evidently the conjugated operators can be defined uniquely,
$\bar K = K^t = K^\dagger, \quad \bar P = P^t = P^\dagger$, independently on
what a choice is taken from \gl{4susy} for the operators $Q, \, \bar Q$.
One can always employ the normalization \gl{superch} of the senior derivative
in $q^\pm_N$ on a real constant. Then the second supercharge $P$ appears to be
a differential operator of lower order $N_1 < N$.

The appearance of the second supercharge conventionally implies the
extension of SUSY algebra. To close the algebra one has to include
all anticommutators between supercharges, {\it i.e.}
the full algebra based on two supercharges $K$ and $P$
with real intertwining operators. Two supercharges generate
two Polynomial SUSY,
\be
\left\{K, K^\dagger\right\} = \tilde{\cal P}_N (H),\quad
\left\{P, P^\dagger \right\} = \tilde{\cal P}_{N_1} (H). \la{2alg}
\ee
This SUSY algebra has to be embedded into a
${\cal N} =2$ SUSY\footnote{There is a misinterpretation concerning the
classification of extended SUSY in QM. The conventional ${\cal N} =1$
SUSY deals with non-hermitian {\it nilpotent} supercharges $Q, \bar Q$ whereas
the ${\cal N} =2$ SUSY should employ two pairs of nilpotent supercharges
$Q_j, \bar Q_j$ satisfying the extended SUSY algebra
with certain central charges.
We are grateful to A.~Smilga for the discussion of this point. However
there are papers (see discussion in \ci{lima})
where the SUSY QM algebra is defined as ${\cal N} =2$ SUSY
in terms of hermitian supercharges $Q_1 = Q + \bar Q,\, Q_2 = i(Q - \bar Q) $.
We would like to stress that an elementary SUSY charge is {\it nilpotent}
carrying fermion quantum numbers. Moreover a nontrivial dynamics cannot be
obtained with one real SUSY charge (as it was recently
mentioned in \ci{lima}).}.
The closure of the extended, ${\cal N} =2$ SUSY algebra is given by
\ba
\left\{P, K^\dagger \right\} & \equiv & {\cal R}
= \left(\begin{array}{cc}
 p^+_{N_1} k^-_{N} &  0\\
0 & k^-_{N} p^+_{N_1}
\end{array}\right),\no
\left\{K, P^\dagger\right\} &\equiv& \bar{\cal R}
= \left(\begin{array}{cc}
 k^+_{N} p^-_{N_1} &  0\\
0 & p^-_{N_1} k^+_{N}
\end{array}\right). \la{roper}
\ea
Evidently the
components of operators ${\cal R},\, \bar{\cal R}= {\cal R}^\dagger ={\cal R}^t$
are differential
operators of $N + N_1$ order commuting with the Hamiltonians $h^\pm$, 
hence they form symmetry operators  ${\cal R},\, \bar{\cal R}$ for the Super-Hamiltonian.
 However, in general, they are not
polynomials of the Hamiltonians $h^\pm$ and these symmetries impose certain
constraints on potentials\footnote{Such type of symmetries in
one-dimensional QM
 and their possible relation to the Lax method in the soliton theory
was discussed in
\ci{matveev,zhdan}.}.

All four operators
$\tilde{\cal P}_N (H),\,  \tilde{\cal P}_{N_1} (H),\, {\cal R},\, \bar{\cal R}$
commute each to other. Moreover the hermitian matrix describing this
 ${\cal N}=2$ SUSY,
\ba
{\cal Z} (H) = \left(\begin{array}{cc}
 \tilde{\cal P}_N (H) & {\cal R}  \\
\bar{\cal R} & \tilde{\cal P}_{N_1} (H)
\end{array}\right), \quad \mbox{\rm det} \left[{\cal Z} (H)\right] =
 \tilde{\cal P}_N \tilde{\cal P}_{N_1} -  {\cal R} \bar{\cal R} = 0, \la{centr}
\ea
is degenerate. Therefore it seems that the two supercharges are not
independent and by their redefinition
(unitary rotation) one might reduce the extended SUSY to an ordinary
${\cal N}=1$ one. However such rotations cannot be global and must
use non-polynomial, pseudo-differential operators as ``parameters''.
Indeed, the operator components of the ``central charge'' matrix ${\cal Z} (H)$
have different order in derivatives. Thus, globally the extended nonlinear
SUSY
deals with two sets of supercharges but when they act on
a given eigenfunction of the
Super-Hamiltonian $H$ one could, in principle, perform the energy-dependent
rotation and
eliminate a pair of supercharges. Nevertheless this reduction can be
possible only after the constraints on potentials have been resolved.

Let us find the formal relation between
the symmetry operators ${\cal R}, \,\bar {\cal R}$ and the Super-Hamiltonian.
These operators can be decomposed into a hermitian and an antihermitian
parts,
\be
{\cal B}\equiv \frac12({\cal R} +\bar {\cal R}) \equiv \left(\begin{array}{cc}
b^+ & 0\\
0 & b^-
\end{array} \right),\quad
i {\cal E} \equiv \frac12 ({\cal R} - \bar {\cal R})\equiv i \left(\begin{array}{cc}
e^+ & 0\\
0 & e^-
\end{array} \right). \la{herm}
\ee
The
operator ${\cal B}$ is a differential operator of even order and therefore
it may be a polynomial of the Super-Hamiltonian $H$. But if the
operator ${\cal E}$ does not vanish identically
it is a differential operator of {\it odd} order and
 cannot be realized by a polynomial of $H$.

The first operator
plays essential role in the one-parameter
non-uniqueness of the SUSY algebra.
Indeed, one can always redefine  the higher-order supercharge as
follows,
\be
 K^{(\zeta)} =  K + \zeta P,\quad
 \left\{ K^{(\zeta)}, K^{(\zeta)\dagger}\right\} =
\tilde{\cal P}^{(\zeta)}_N (H), \la{redef}
\ee
keeping the same order $N$ of Polynomial SUSY for arbitrary real
parameter $\zeta$. From \gl{redef} one gets,
\be
2 \zeta  {\cal B} (H) =
\tilde{\cal P}^{(\zeta)}_N (H) -\tilde{\cal P}_N (H) - \zeta^2
 \tilde{\cal P}_{N_1} (H),
\ee
thereby the hermitian operator ${\cal B}$ is a
polynomial of the Super-Hamiltonian of the order $N_b \leq N -1$.
Let's use it to unravel the Super-Hamiltonian content of the operator ${\cal E}$,
\be
{\cal E}^2 (H) = \tilde{\cal P}_N (H)
\tilde{\cal P}_{N_1} (H) - {\cal B}^2 (H), \la{secsym}
\ee
which follows directly from \gl{centr} and \gl{herm}. As
the (nontrivial) operator ${\cal E} (H)$
is a differential operator of odd order $N_e$ it may have only a
realization non-polynomial in $H$ being a square root of \gl{secsym} in
an operator sense. This operator is certainly non-trivial if the sum of orders
$N + N_1$ of the operators $k^\pm_N$ and $p^\pm_{N_1}$ is odd and therefore
$N_e = N + N_1$. For an even sum $N + N_1$ we cannot in general
make any definite conclusion concerning the non-triviality
of  ${\cal E} (H)$. However  it will be shown in Sec.~7 that if the symmetry
operator is non-zero then
for any choice of the operators $k^\pm_N$ and $p^\pm_{N_1}$ an optimal set of
independent supercharges (possibly of lower orders)
can be obtained which is
characterized by an odd sum of their orders.

The existence of a
nontrivial symmetry operator ${\cal E}$ commuting with the
Super-Hamiltonian results in common eigenstates which however are not necessarily
physical wave functions. In general they may be combinations of two solutions
of the Shr\"odinger equation with a given energy, the physical and
unphysical ones. But if  the
symmetry operator ${\cal E}$  is
hermitian in respect to the scalar product of the
Hilbert space spanned on the eigenfunctions of the Super-Hamiltonian $H$ then
both operators have a common set of physical wave functions. This fact imposes quite rigid
conditions on  potentials.

In particular,
for
intertwining operators with sufficiently smooth coefficient functions having
constant asymptotics at large coordinates the
symmetry operator ${\cal E}$  has the similar properties and is evidently
hermitian. In this case one has  non-singular
potentials with constant asymptotics at large $x$ and therefore
a continuum energy spectrum  of $H$
with wave functions satisfying the scattering conditions.
Thus the incoming and outgoing states, $\psi_{in}(x)$ and $\psi_{out}(x)$,
at large $x$ are conventionally represented by  combinations of plane waves
which are solutions of the Schr\"odinger equation for a free particle,
\ba
&&\psi (x)|_{x \rightarrow -\infty} \longrightarrow
\exp(ik_{in}x) + R(k_{in})\exp(-ik_{in}x),\no
&&\psi(x)|_{x \rightarrow +\infty} \longrightarrow
\left(1 + T(k_{out})\right)\exp(ik_{out}x),
\ea
where the reflection, $R(k_{in})$, and transmission, $T(k_{out})$, coefficients
are introduced. Since the symmetry  is described by a differential operator
of odd order  which at large $x$ tends to an antisymmetric
operator with constant
coefficients the eigenstates of this operator at large coordinates
approach to individual plane waves
$\sim \exp(\pm ikx)$ with opposite eigenvalues $\sim \pm k f(k^2)$
and cannot be their combinations. Hence the eigenstate of the
Super-Hamiltonian with a given value of the operator ${\cal E}$ may characterize
only the transmission and cannot have any reflection, $R(k_{in}) = 0$.
We conclude that the corresponding potentials $V_{1,2}$
in \gl{suham} inevitably belong to the
class of transparent or reflectionless ones \ci{refl}.

As the
symmetry operator ${\cal E}$  is
hermitian its eigenvalues are real but, by construction,
its coefficients are purely imaginary. Since the
wave functions of bound
states of the system $H$ can be always chosen real we conclude that they
must be zero-modes of the operator ${\cal E} (H)$,
\be
{\cal E} (H) \psi_i = {\cal E} (E_i) \psi_i = 0,\quad
  \tilde{\cal P}_N (E_i)\tilde{\cal P}_{N_1} (E_i) - {\cal B}^2 (E_i) = 0,
\la{zeroeq}
\ee
which represents the algebraic equation on bound state
energies of a system possessing two supersymmetries. Among solutions
of \gl{zeroeq} one reveals
also a zero-energy state at the bottom of
continuum spectrum. On the other hand
one could find also the solutions which are not
associated to any bound state. The very
appearance of such unphysical solutions is
accounted for by the trivial possibility to replicate supercharges by their
multiplication on the polynomials of the Super-Hamiltonian and it is discussed
in Sec.~6.

\section{Example: $N = 2, N_1 = 1$}
\hspace*{2ex} Let us examine the algebraic structure of the simplest non-linear SUSY
with two supercharges,
\ba
k^\pm &\equiv & \partial^2 \mp 2f(x)\partial + \tilde b(x) \mp f'(x) ; \no
p^\pm &\equiv & \mp\partial + \chi(x), \la{exgen}
\ea
induced by the complex supercharge
of second order in derivatives \ci{comalg}.
The supersymmetries \gl{extsusy}
generated by $K, \bar K$ and $P, \bar P$ prescribe that
\ba
V_{1,2} &=& \chi^2 \mp \chi' =
\mp 2f' + f^2 + \frac{f''}{2f} - \left(\frac{f'}{2f}\right)^2 -
\frac{d}{4f^2} -a,\no
\tilde b &=& f^2 - \frac{f''}{2f} +\left(\frac{f'}{2f}\right)^2 +
\frac{d}{4f^2},
\ea
where $\chi, f$ are real functions and $a, d$ are real constants.
The related superalgebra closure for $K, \bar K$ and $P, \bar P$
takes the form,
\be
\{K, \bar K\} = (H + a)^2 + d,\quad  \{P, \bar P\} = H,
\la{secor}
\ee
the latter one clarifies the role of constants $a, d$.

The compatibility of two supersymmetries is achieved on solutions
of the following equations
\be
\chi = 2 f + \chi_0,\quad f^2 + \frac{f''}{2f} -
\left(\frac{f'}{2f}\right)^2 - \frac{d}{4f^2} -a = \chi^2 =
 (2 f + \chi_0)^2, \la{2equ}
\ee
where $\chi_0$ is an arbitrary real constant.
The latter one represents a nonlinear second-order differential equation
which solutions are parameterized by two integration constants. Therefore
as it was advertised the existence of two SUSY constrains substantially
the class of potentials for which they may hold.

Let us use the freedom to redefine the higher-order supercharge \gl{redef}
for eliminating the constant $\chi_0$ in \gl{2equ}. After this
simplification
the equation \gl{2equ} is  integrated into the following, first-order one,
\be
\chi = 2 f;\quad
(f')^2 = 4 f^4 + 4 a f^2 + 4 G_0 f - d \equiv \Phi_4(f),
\la{firstor}
\ee
where $G_0$ is a real constant.

The solutions of this equation are elliptic functions which
 can be easily found in the implicit form,
\be
\int^{f(x)}_{f_0} \frac{df}{\sqrt{\Phi_4(f)}} = \pm (x - x_0),
\ee
where the lower limit of integration $f_0$ and $x_0$ are real constants.

It can be shown that they may be nonsingular in three situations.\\
a) The polynomial $\Phi_4(f)$ has four different real roots $f_1 \leq f_2 \leq f_3 \leq f_4$ and
$f_0$ is chosen between two roots $f_2$ and $f_3$.
The corresponding potentials are periodic. This case will not be
examined here.\\
b)  $\Phi_4(f)$ has three different real roots and the double root
$\beta/2$
is either the maximal one or a minimal one,
\be
 \Phi_4(f) = 4 (f - \frac{\beta}{2})^2
\left((f + \frac{\beta}{2})^2 - (\beta^2 -
\epsilon)\right),\quad
 0 < \epsilon < \beta^2. \la{polyn}
\ee
Then there exits a relation
between constants $a, d, G_0$ in terms of
coefficients $\beta,\epsilon$,
\be
a = \epsilon - {3 \beta^2\over 2} < 0,\quad G_0 = \beta
(\beta^2 - \epsilon),
\quad d = \beta^2  \left({3 \beta^2\over 4} - \epsilon \right) .
\la{defa}
\ee
Besides the constant $f_0$ is
taken between the double root and a nearest simple root.\\
c)  $\Phi_4(f)$ has two different real double roots which corresponds
in \gl{polyn}, \gl{defa} to
$G_0 = 0,\quad \beta^2 = \epsilon > 0, \quad a = -\epsilon/2,\quad
d = -\epsilon^2/4$. The constant $f_0$ is
taken between the roots.

The corresponding potentials $V_{1,2}$ are well known \ci{refl} and in the
cases b) and c) are
reflectionless, with one bound state at the energy $ (\beta^2 -
\epsilon)$ and
with the continuum spectrum starting from $ \beta^2$.
Respectively the scattering wave function is proportional to $\exp(ikx)$
with $k = \sqrt{E - \beta^2 }$.

In particular, in the case b) the potentials coincide in their form
and differ only  by shift in the coordinate,
\be
V_{1,2} =  \beta^2 -
\frac{2\epsilon }{\mbox{\rm ch}^2 \left(\sqrt{\epsilon}(x -
x^{(1,2)}_0)\right)},\quad x^{(1,2)} =x_0 \pm \frac{1}{4\sqrt{\epsilon}}
\ln\frac{\beta - \sqrt{\epsilon}}{\beta + \sqrt{\epsilon}},
\la{caseb}
\ee
and  in the case c) one of the potentials can be chosen constant,
\be
V_1 = \beta^2,\quad V_2 =  \beta^2 \left(1 -
\frac{2}{\mbox{\rm ch}^2 \left(\beta (x - x_0)\right)}\right),
\la{casec}
\ee
For these potentials one can illustrate
all the relations of extended SUSY algebra.

The initial algebra is given by the
relations \gl{secor}. The first, polynomial
symmetry operator turns out to be constant, $ {\cal B} (H) = G_0$
when taking into account \gl{exgen} and \gl{firstor}. The second
symmetry operator reads,
\be
{\cal E} (H) = i\left[{\bf I}\,\partial^3 - \Bigl(a {\bf I} + 
\frac32 {\bf V}(x)\Bigr) \partial 
- \frac34 {\bf V}' (x)\right],
\la{oddrep}
\ee
in terms of the potential \gl{suham}. From the identity \gl{secsym} or
directly from Eq.~\gl{oddrep}
one derives with the help of Eqs.~\gl{firstor} and \gl{defa} that,
\be
{\cal E}^2 (H) = H \left[ (H + a)^2 + d\right]
 -  G_0^2  =  (H - E_b)^2 (H -  \beta^2),\la{quadr}
\ee
where $E_b = \beta^2 - \epsilon $ is the energy of a bound state.
Thus (some of) the zero modes of ${\cal E} (H)$ characterize either bound states or
zero-energy states in the continuum. However there exist also the 
non-normalizable, unphysical zero-modes corresponding to $E = E_b, \beta^2$. 
We remark that in the case c)
only the Hamiltonian $h^-$ has a bound state. Hence the physical 
zero modes of
 ${\cal E} (H)$ may be either degenerate (case b), broken SUSY) or (one
of them) non-degenerate (case c), unbroken SUSY).

The square root in \gl{quadr} can be carried out,
\be
{\cal E} (H) =  (H - E_b) \sqrt{H -  \beta^2}.\la{root}
\ee
We notice that the symmetry operator \gl{oddrep}, \gl{root} is irreducible, {\it i.e.}
 the binomial $ (H - E_b)$ cannot be ``stripped off''
(the exact meaning of this operation see in Sec.~6). Indeed the
elimination of this binomial would lead to an essentially nonlocal
operator. The sign of square root in \gl{root}
is fixed from the conventional asymptotics
of scattering wave functions
$\sim \exp(ikx)$ and the asymptotics $V_{1,2} \longrightarrow \beta^2$
by comparison of this relation with Eq.~\gl{oddrep}.

When taking Eq.~\gl{root} into account one finds the non-polynomial relations
of the extended  SUSY algebra,
\be
\{ K, P^\dagger\} = \{K^\dagger, P\}^\dagger =
G_0 - i (H - E_b) \sqrt{H -  \beta^2}. \la{nonpol}
\ee
The hermitian matrix ${\cal Z} (H)$, Eq.~\gl{centr} is built of the elements
\gl{secor} and \gl{nonpol} and evidently cannot be diagonalized by a unitary
rotation with elements polynomial in $H$. Thus the algebra must be
considered to be extended in the class of
differential operators of finite order.

It remains to clarify the very non-uniqueness of the
higher-order supercharge, namely, its role in the classification of the
Polynomial SUSY. For arbitrary $\zeta$ in \gl{redef} one obtains
\ba
&&\{K^{(\zeta)}, K^{(\zeta)\dagger}\} = H^2 + (2a + \zeta^2) H + a^2 + d
+ 2 \zeta G_0 = (H + a_\zeta)^2  + d_\zeta,\no
&& a_\zeta = a + \frac12 \zeta^2,\quad
d_\zeta = d + 2 \zeta G_0 - a \zeta^2 - \frac14 \zeta^4
\equiv - \Phi_4 (- \frac{\zeta}{2}),
\ea
where $ \Phi_4 (f)$ is defined in Eq.~\gl{firstor}.

One can see that the sign of $d_\zeta$, in general, depends on the choice of $\zeta$.  For instance, let us consider the case b) when
\be
d_\zeta  = - \frac14 \left(\zeta + \beta\right)^2
\left[ \left(\zeta - \beta \right)^2 - 4 (\beta^2 -\epsilon)\right].
\la{dlam}
\ee
Evidently if $\zeta$ lies in between the real
roots of the last factor in \gl{dlam}
then $d_\zeta$ is positive and otherwise it is negative. But two real roots
always exist because $\beta^2 >\epsilon$.
Thereby the sign of $d_\zeta$
can be freely negative or positive without
any change in the Hamiltonians. Hence  in the case when
the Polynomial SUSY is an extended one,
with two sets of supercharges, the irreducibility or
reducibility of a Polynomial SUSY algebra does not signify any
invariant characteristic of potentials.

\section{Complex SUSY algebras}
\hspace*{2ex}If the intertwining
operators $q^\pm$ have complex coefficients in
\gl{superch} then  we deal with two supercharges which we
adopt to be independent (see Sec.~7 for its exact definition).
One can split again the
complex supercharge $Q$ into a real, $K$ and an imaginary, $P$ counterparts
as in Eq.~\gl{extsusy} and normalize them so that the intertwining operator
in $K$ has a higher order in derivatives. Two SUSY algebras
with transposition symmetry,
${\cal A}_2$ and ${\cal A}_3$, are polynomial in virtue of the Theorem
of Sec.~2. In terms of real supercharges they have the following structure,
\ba
&&\{Q, \bar Q_c\} =  \tilde{\cal P}_N (H)
- \tilde{\cal P}_{N_1} (H) + i 2{\cal B} (H),\la{2susy1}\\
&&\{Q_c, \bar Q\} =  \tilde{\cal P}_N (H)
- \tilde{\cal P}_{N_1} (H) -i 2 {\cal B} (H), \la{2susy2}\\
&&\bar Q_c = Q^t =
K^\dagger + i P^\dagger;\quad  Q_c = Q^* =
K - i P;\quad \bar Q =
K^\dagger - i P^\dagger. \la{2def}
\ea
Two more algebras, ${\cal A}_1$ and ${\cal A}_4$ can be built
with a hermitian closure according to Eq.~\gl{4susy}. In particular,
the algebra  ${\cal A}_1$ (used in the
AST theorem, Sec.~1) is completed by the following closure,
\be
\{Q, \bar Q\} =  \tilde {\cal P}_N (H)
+ \tilde{\cal P}_{N_1} (H) - 2{\cal E} (H). \la{2susy3}
\ee
Respectively the algebra ${\cal A}_4$ is completed by the relation,
\be
\{Q_c, \bar Q_c\} =  \tilde{\cal P}_N (H)
+ \tilde{\cal P}_{N_1} (H)  + 2{\cal E} (H). \la{2susy4}
\ee
When the symmetry operator ${\cal E} (H)$ is nontrivial
they are essentially non-polynomial (see \gl{secsym}).

We conclude that for complex intertwining operators
the same pair of isospectral Hamiltonians may be induced
both by the polynomial SUSY algebra \gl{2alg} (or
\gl{2susy1}, \gl{2susy2}) and by the non-polynomial one
\gl{2susy3}, \gl{2susy4}.

All four superalgebras ${\cal A}_m$ generated by
$(Q_1, Q_2) \equiv (Q, Q_c)$
can be assembled into the extended ${\cal N} = 2$ SUSY
algebra,
\be
\{Q_i, \bar Q_j\} = \left[({\bf I}+ \tau_1) \tilde{\cal P}_N (H)
+
({\bf I}- \tau_1) \tilde{\cal P}_{N_1} (H) - \tau_2 2{\cal B} (H)
- \tau_3 2 {\cal E} (H)\right]_{ij},
\ee
with $ i,j = 1,2$. It is equivalent, of course,
to the algebra \gl{2alg} and \gl{roper}.

Let us illustrate such an algebra using the  $N = 2, N_1 = 1$ example of
Sec.~4. Thus the intertwining operator $q^+ = k^+ + ic p^+$ is composed from
the operators \gl{exgen} where the constant $c$ of mass dimension
is introduced from dimensional reasons. Respectively,
\ba
\{Q_i, \bar Q_j\} &=& \left[({\bf I}+ \tau_1) \left((H + a)^2 + d \right)
+({\bf I}- \tau_1) c^2 H\right. \no
&&\left. - \tau_2\, 2c G_0 -
\tau_3\, 2c (H - E_b) \sqrt{H -  \beta^2}\right]_{ij},
\ea
i.e. is manifestly non-polynomial in respect to the Super-Hamiltonian $H$.

It remains to clarify the relationship between the hermitian algebra
 ${\cal A}_1$ determined in the AST theorem by Eq.~\gl{ASTrep}
and that one given by \gl{2susy3}. For this purpose
we relate the representation \gl{ASTrep}
to the algebra with transposition symmetry,
Eq.~\gl{tposusy}. In order to establish the exact correspondence
the upper and lower components in the matrix $\{Q, \bar Q\}$ have to
be treated differently.
Namely for the upper component $q^+_N q^-_N $ the suitable decomposition,
$q^+_N = (q^-_N)^t + 2i p^+_{N_1}$, yields,
\be
q^+_N q^-_N =(q^-_N)^t q^-_N +  2i p^+_{N_1}q^-_N =
{\rm det} \left[E{\bf I}-\bar{\bf S}^{+}\right]_{E = h^+} +  2i p^+_{N_1}q^-_N,
\la{ASTrec1}
\ee
where from one reproduces the upper component of Eq.~\gl{ASTrep} after
the identification $ \pi^+_1 \equiv 2i p^+_{N_1}$. We stress that the matrix
$\bar{\bf S}^{+}$ does not coincide with ${\bf S}^{+}$ from
Eq.~\gl{quasiham1} because in \gl{ASTrec1} $q^-_N = (q^+_N)^\dagger
\neq (q^+_N)^t$. But in appropriate bases of mutually complex conjugated
functions
they may be related by complex conjugation, $\bar{\bf S}^{+} =
({\bf S}^{+})^*$.

Similarly the lower component can be transformed into the form \gl{ASTrep}
by means of the decomposition, $q^-_N = (q^+_N)^t - 2i p^-_{N_1}$ and reads,
\be
q^-_N q^+_N =(q^+_N)^t q^+_N -  2i p^-_{N_1}q^+_N =
{\rm det} \left[E{\bf I}- \bar {\bf S}^{-}\right]_{E = h^-} -  2i p^-_{N_1}q^+_N,
\la{ASTrec2}
\ee
where from one obtains the lower component of Eq.~\gl{ASTrep} after
the identification $ \pi^-_2 \equiv -2i p^-_{N_1} = \pi^-_1 $.
The matrix $\bar{\bf S}^{-}$
is exactly as ${\bf S}^{-}$
in the Theorem on SUSY algebras with T-symmetry,
Eq.~\gl{quasiham1}.

The AST decomposition \gl{ASTrec1}, \gl{ASTrec2} is certainly equivalent
to the representation \gl{2susy3} but supplies both the polynomial part
and the non-polynomial symmetry operator with  imaginary contributions  which
eventually assemble into  $-2e^\pm(h^\pm)$. Thereby the hermitian
symmetry operator ${\cal E} (H)$ non-polynomial in $H$ is hidden in
Eqs.~\gl{ASTrec1}, \gl{ASTrec2}.  This is why we give our favor to the
representation \gl{2susy3} in the analysis of supersymmetries with several
supercharges.

\section{``Strip-off'' problem}
\hspace*{2ex} The pair of two
supersymmetries analyzed in Sec.~3 and 5
may  rigidly determine the class of potentials $V_{1,2}$
in \gl{suham}  to a specific, transparent type of them
contracting the freedom in their
choice from a functional one to a parametric one.  On the other hand,
there exists a trivial possibility when the intertwining operators
$k^\pm_N$ and
$p^\pm_{N_1}$ are related by a factor depending on
the Hamiltonian,
\be
k^\pm_N =
F (h^\pm) p^\pm_{N_1}
=  p^\pm_{N_1}  F (h^\mp), \la{triv}
\ee
where
$F(x)$ is assumed to be a polynomial. Obviously in
this case
the symmetry operator ${\cal E} (H)$ identically vanishes and the appearance
of the second supercharge does not result in any restrictions on potentials.

More generally the orders of  polynomial superalgebras and some of
the roots of
associated polynomials may not be involved in determination of
the structure of the potentials.
In particular, let the operators $k^\pm_N$ and
$p^\pm_{N_1}$  be reducible
to some lower-order ones $\tilde k^\pm_{\tilde N}$ and
$\tilde p^\pm_{\tilde N_1}$,
\be
k^\pm_N =
F_k (h^\pm) \tilde k^\pm_{\tilde N}
=  \tilde k^\pm_{\tilde N}  F_k (h^\mp),\qquad
 p^\pm_{N_1} =   F_p (h^\pm) \tilde p^\pm_{\tilde N_1}
=  \tilde p^\pm_{\tilde N_1} F_p (h^\mp), \la{triv1}
\ee
where the numbers in the pairs $N,\tilde N$ and $N_1, \tilde  N_1$
are simultaneously
odd or even and
$F_k(x), F_p(x)$ are  polynomials of order
$(N - \tilde N)/2$ and $(N_1 - \tilde N_1)/2$.
Then evidently the superalgebra generated by
 $\tilde k^\pm_{\tilde N}$ and
$\tilde p^\pm_{\tilde N_1}$ equally well characterizes the Super-Hamiltonian system
with the same potentials.

We have come to the problem of how to discern the nontrivial part of a
supercharge and avoid multiple SUSY algebras generated by means of ``dressing''
\gl{triv1}. It can be systematically performed with the help of the
following theorem.\\

\noindent
\underline{``Strip-off'' Theorem }\\
{\it
Let's admit the
construction of the Theorem on SUSY algebras with T-symmetry. Then the
requirement\\
that
the matrix $\widetilde{\bf S}^-$ (or $\widetilde{\bf S}^+$) generated on the
subspace of zero-modes of the operator $k^+_N$ (or $k^-_N$)
contains $n$ pairs (and no more) of Jordan
cells with equal
eigenvalues  $\lambda_l$
and the sizes $\nu_l$ of a smallest cell in the $l$-th pair\\
is necessary and sufficient to ensure for
the intertwining operator $k^+_N$ (or $k^-_N$) to be represented in the
factorized form:
\be
k^\pm_{N} = \tilde k^\pm_{\tilde N} \prod^n_{l=1} (\lambda_l - h^\mp)^{\nu_l}, \la{factor}
\ee
where $ \tilde k^\pm_{\tilde N}$
are intertwining operators\footnote{\rm In this theorem the
intertwining operators $k^\pm_{N},  \tilde k^\pm_{\tilde N}$ and the parameters
$\lambda_l$ may also be complex.}
of order $\tilde N = N - 2 \sum^n_{l=1}\nu_l $
which cannot be  decomposed
further on in the product similar to \gl{triv1} with $F_k (x) \not= const$.}

\medskip

We shall perform the {\it proof} of the theorem for
$\widetilde{\bf S}^-$ only as its proof for $\widetilde{\bf S}^+$ is similar.
It is based on the lemma and two remarks.\\

\nn
\underline{Remark 1.} The matrices  $\widetilde{\bf S}^\pm$ cannot contain more
than two Jordan cells with the same eigenvalue $\lambda$ because otherwise the
operator $\lambda - h^\pm$ would have more than two linearly independent
zero-modes (not necessarily normalizable).\\

\noindent
\underline{Lemma.}\\
{\it
In order that the intertwining operator $k^+_N$ could be factorized,
\be
k^+_N = k^+_{N-2} (\lambda - h^-), \la{factor2}
\ee
with $k^+_{N-2}$  being another intertwining operator of order $N-2$,
it is necessary and sufficient for the matrix
$\widetilde{\bf S}^-$ to contain two Jordan cells with the same eigenvalue
$\lambda$.}\\

\noindent
\underline{\it Proof}.\hspace{3ex} The requirement of this Lemma is
sufficient because if  $\widetilde{\bf S}^-$ contains
two Jordan cells with the same eigenvalue $\lambda$ then the kernel of $k^+_N$
includes two linearly independent solutions of the equation
$h^-\phi = \lambda \phi$.  When repeating the way of proof of the Theorem
on SUSY algebras with T-symmetry one can derive that $k^+_N$ is factorized
in the form,
\be
k^+_N = k^+_{N-2} (\partial - \chi_2(x))(\partial - \chi_1(x)),
\ee
where  $k^+_{N-2}$  is a differential operator of order $N-2$ and the functions
$\chi_1(x)$ and $\chi_2(x)$ are chosen to provide the equal kernels of
operators $\lambda - h^-$ and $(\partial - \chi_2)(\partial - \chi_1)$.
As a differential operator of second order with the unit coefficient at
$\partial^2$ is uniquely determined by two linearly independent elements
of its kernel one concludes that
\be
\lambda - h^- = (\partial - \chi_2)(\partial - \chi_1),
\ee
and therefore \gl{factor2} is valid. At last, from the relations,
\be
k^+_{N-2} h^-(\lambda - h^-) = k^+_N h^- = h^+ k^+_N =  h^+ k^+_{N-2}(\lambda - h^-),
\ee
one obtains that the operator $k^+_{N-2}$ is intertwining.

The condition \gl{factor2} is also necessary as in this case the
kernel of $k^+_N$ includes two linearly independent solutions of the
equation $h^-\phi = \lambda \phi$ which induce two
different Jordan cells with the same eigenvalue.\\

\medskip

\nn
\underline{Remark 2.} Within the Lemma let us put
the lower lines of two Jordan cells with
the eigenvalue $\lambda$ into the $i$-th and $j$-th line ($i < j$) of
$\widetilde{\bf S}^-$ respectively and introduce the functions,
\ba
\tilde \psi^+_l(x) =\left\{\begin{array}{cc}
(\lambda - h^-)\tilde\phi^+_l,& 1\leq l \leq i-1,\\
(\lambda - h^-)\tilde\phi^+_{l+1},& i\leq l \leq j-2,\\
(\lambda - h^-)\tilde\phi^+_{l+2},& j-1\leq l \leq N-2,
\end{array}\right.
\ea
where the functions $\tilde\phi^+_m$ form the basis of the kernel of
the operator  $k^+_N$ which supply the matrix $\widetilde{\bf S}^-$ with
the Jordan form.
Evidently the functions $\tilde \psi^+_l(x)$ are linearly independent because
in the opposite case a nontrivial linear combination of
$\tilde\phi^+_1,\cdots\tilde\phi^+_{i-1},$
$\tilde\phi^+_{i+1},\cdots \tilde\phi^+_{j-1},$
$\tilde\phi^+_{j+1},\cdots \tilde\phi^+_N$ must be a combination of
zero-modes $\tilde\phi^+_{i}$ and $\tilde\phi^+_{j}$. Thus they
form the complete set of solutions of the equation $ k^+_{N-2}\tilde\psi = 0$
which
comes from \gl{factor2}. The matrix $\widetilde{\bf S}^-_{N-2}$ which
describes the Hamiltonian action on the zero-mode subspace
$\{\tilde \psi^+_k(x)\}$ of the intertwining operator $k^+_{N-2}$,
can be produced from the matrix $\widetilde{\bf S}^-$ by
deleting both the $i$-th and $j$-th columns and lines. Thereby it has a
Jordan form.\\

Now from the Lemma and the Remarks 1 and 2 one derives all
statements of the ``Strip-off'' Theorem.

This theorem naturally supplements the Theorem on SUSY algebras with
T-symmetry as it entails the essential identity of the Jordan forms
 $\widetilde{\bf S}^-$ and  $\widetilde{\bf S}^+$ (up to transposition of
certain Jordan cells).

Let us  illustrate the  Theorem by the \underline{Example:}\,
the matrix $\widetilde{\bf S}^-$ for the intertwining operator $k^+_3$
with Jordan cells of different size having
the same eigenvalue. It is generated by the operators,
\be
p^\pm = \mp\partial + \chi,\quad h^\pm =p^\pm p^\mp + \lambda,\quad
k^+_3 = - p^+ p^- p^+ =p^+ (\lambda - h^-).
\ee
Respectively:
\be
\begin{array}{cccc}
\tilde\phi^+_2:\,& p^+\tilde\phi^+_2 = 0 &\longrightarrow& h^- \tilde\phi^+_2 = \lambda \tilde\phi^+_2;\\
\tilde\phi^+_3:\,& p^+\tilde\phi^+_3 \not= 0,\, p^- p^+\tilde\phi^+_3 = 0&
\longrightarrow &h^- \tilde\phi^+_3 = \lambda \tilde\phi^+_3;\\
\tilde\phi^+_1:\,& p^- p^+\tilde\phi^+_1 = \tilde\phi^+_2 &\longrightarrow&
h^- \tilde\phi^+_1 = \lambda \tilde\phi^+_1 + \tilde\phi^+_2.
\end{array}
\ee
Thus,
\be
\widetilde{\bf S}^- =\left(\begin{array}{ccc}
\lambda&1&0\\
0&\lambda&0\\
0&0&\lambda
\end{array}\right).
\ee

As a consequence of the ``Strip-off'' Theorem one finds that
the supercharge components cannot be factorized
in the form \gl{triv1}
if the polynomial $\tilde{\cal P}_N (x)$ in the SUSY algebra closure \gl{2alg}
does not reveal the degenerate zeroes. Indeed the SUSY algebra closure
contains the square of polynomial $F(x)$, for instance,
\be
k^-_N k^+_N = F_k (h^-) \tilde k^-_{\tilde N} \tilde k^+_{\tilde N}  F(h^-)
=  F_k^2 (h^-) \tilde{\cal P}_{\tilde N} (h^-),
\la{dzero}
\ee
where $\tilde{\cal P}_{\tilde N} (x)$ is a polynomial of
lower order, $\tilde N \leq N-2$.
Therefore each zero of the polynomial $F_k (x)$ will produce a double zero
in the SUSY algebra provided by \gl{dzero}.

Thus the absence of double zeroes
is sufficient to deal with the SUSY charges non-factorizable in the sense
of Eq.~\gl{triv1}. However it is not necessary because the degenerate zeroes
may well arise in the ladder (dressing chain) construction giving
new pairs of isospectral potentials (see, for instance, \ci{acdi} for the
Polynomial SUSY of second order).

Further on we consider only the stripped-off supercharges. In this case
the existence of two intertwining operators results in more equations
on their coefficient functions and thereby on the potentials in $h^\pm$.
When they are compatible they rigidly dictate the form of potentials
leaving only the parametric freedom for their choice.

Still the stripped-off supercharges do not necessarily represent an optimal 
set of them and provide an optimal
structure of the symmetry operator ${\cal E} (H)$.

Let us illustrate it with
the sample intertwining operators $ t^\pm =p_{N_1}^\pm k_N^\mp p_{N_1}^\pm$
made of two stripped-off supercharges  with components $k_N^\pm$ and $p_{N_1}^\pm$. If
the differential operators $k_N^\pm$ and $p_{N_1}^\pm$ have even and odd order
respectively and the operators $b^\pm$ are non-zero the operators
$t^\pm$ cannot be stripped off till  $k_N^\pm$ or $p_{N_1}^\pm$ because
the components of symmetry operator $e^\pm = \pm \frac{i}{2}
(k_N^\pm p_{N_1}^\mp - p_{N_1}^\pm k_N^\mp)$
are surely non-trivial and therefore the operators
$k_N^\pm p_{N_1}^\mp$ and $p_{N_1}^\pm k_N^\mp$ are 
not polynomials of the Hamiltonian. One can see it manifestly for the 
supercharges \gl{exgen} of the Example in the case b) when $G_0 \not=0$.

Meantime the system
composed of two supercharges
with components $t^\pm$ and $p_{N_1}^\pm$ has
the symmetry realized by the operator
 ${\cal E}_t (H)$  with components,
\be
e^\pm_t =
\pm i\frac12 (t^\pm p_{N_1}^\mp - p_{N_1}^\pm t^\mp) =
\mp i\frac12 (k_N^\pm p_{N_1}^\mp - p_{N_1}^\pm k_N^\mp)
p_{N_1}^\pm p_{N_1}^\mp = - e^\pm \tilde{\cal P}_{N_1} (h^\pm),
\ee
where to obtain it
the commutation of the Super-Hamiltonian
with operators ${\cal R},\, \bar{\cal R}$,
Eq.~\gl{roper} has been used. Thus the symmetry operator  ${\cal E}_t (H)$
contains a  polynomial factor
$\tilde{\cal P}_{N_1} (H)$ which zeroes are not in general related to any bound state
energies. The lesson is that the symmetry operator must be stripped-off in
addition to the intertwining operators in order
to avoid unphysical zeroes of the function ${\cal E} (E)$. In order to perform the minimization of the
symmetry operator one can employ the "Strip-off" Theorem and analyze the
Jordan form of the relevant Hamiltonian projection $\tilde{\bf S}^\pm_e$ on the
zero-mode subspace of the operator $e^\pm$. Then the elimination of pairs
of equal eigenvalues from different Jordan cells would essentially reduce the
spectrum of the projected Hamiltonian towards a set of its bound state
energies.

On the other hand the intertwining operators $t^\pm$ represent a linear combination of
 $k_N^\pm$ and $p_{N_1}^\pm$ with coefficients depending on the Hamiltonian,
\be
 t^\pm = 2 b^\pm (h^\pm) p_{N_1}^\pm - \tilde{\cal P}_{N_1}(h^\pm)  k_N^\pm.
\la{lincom}
\ee
If the polynomials  $b^\pm$  and $\tilde{\cal P}_{N_1}$ do not have common
roots one may find the intertwining operators  $t^\pm$ which cannot be
stripped off till a combination of  $k_N^\pm$ and $p_{N_1}^\pm$ of lower order in
derivatives (see again the case b) in Sec.~4).

However one can easily build the equivalent supercharges
 $\tilde t^\pm= t^\pm - 2 b^\pm (h^\pm) p_{N_1}^\pm$ which can be stripped off
till  $k_N^\pm$. Thus in order to construct the optimal basis of supercharges
one should not only factorize out the polynomials of the Super-Hamiltonian but also
examine their various linear combinations with coefficients polynomial in
the Super-Hamiltonian.

\section{Optimization of supercharges}
\hspace*{2ex} As it follows from the previous discussion the
existence of several supercharges is controlled by a non-trivial
symmetry operator. If there are several SUSY generators the necessity arises:\\
a) to introduce the notion of (in)dependence of intertwining operators;\quad
b) to find out how many independent supercharges can coexist;\quad
c) to define an optimal basis of intertwining operators.

Let us extrapolate the relation \gl{lincom} and define the intertwining 
operators
$q^\pm_i,\, i=1,\ldots n$ to be dependent if and only if the
polynomials $\alpha_i^\pm (y)$
exist such that not all of them are vanishing and
\be
\sum_{i=1}^n \alpha_i^\pm(h^\pm)q_i^\pm=0. \la{depen}
\ee
If the relation \gl{depen} results in $\alpha_i^\pm (y) = 0$ 
for all $i$  the corresponding SUSY generators are independent.
Evidently the (in)dependence of $q^+_i$ entails  the (in)dependence
of $q^-_i$ and {\it vice versa}.

The following theorem plays a key role in resolution of how many
independent supercharges can commute with a given Super-Hamiltonian.\\

\noindent
\underline{Theorem on (in)dependence of supercharges}\\
{\it
Consider two non-trivial intertwining operators $q^\pm_i,\, i=1,2$ with
transposition symmetry $q^+_i = (q^-_i)^t$ which in general
may have complex coefficients and let us normalize them in accordance to
\gl{superch}. Then the stripped-off intertwining operators
$\tilde q^\pm_i$ coincide if and only if
the symmetry operator made of  $q^\pm_i$ vanishes,
$q^+_1 q^-_2- q^+_2 q^-_1 = 0$ (or equivalently
$q^-_1 q^+_2- q^-_2 q^+_1 = 0$).}

\medskip

The {\it proof} of this theorem is based on the Lemma.\\

\noindent
\underline{Lemma}\\
{\it
If $q^+_1 q^-_2- q^+_2 q^-_1 = 0$  (or equivalently
$q^-_1 q^+_2- q^-_2 q^+_1 = 0$) then the sets of Jordan cells
(and their sizes) in the matrices $\tilde{\bf S}^\pm_{1}$ and
$\tilde{\bf S}^\pm_{2}$
for the operators $\tilde q^\mp_1$ and  $\tilde q^\mp_2$ are identical.}

\medskip

To prove its validity  we first remark that evidently the symmetry operator
 $\tilde q^+_1 \tilde q^-_2 - \tilde q^+_2 \tilde q^-_1$
for stripped-off operators  $\tilde q^\mp_1$ and  $\tilde q^\mp_2$ also
vanishes in virtue of intertwining relations and because the factor
polynomials for $q^+_i$ and $q^-_i$ are equal. Therefore the operator
\be
b^+_{12} = \tilde q^+_1 \tilde q^-_2=  \tilde q^+_2 \tilde q^-_1 =
\frac12 [(\tilde q^+_1 +\tilde q^+_2)(\tilde q^-_1 + \tilde q^-_2) -
\tilde q^+_1 \tilde q^-_1 - \tilde q^+_2\tilde q^-_2]
\ee
is a polynomial of the Hamiltonian $h^+$ (compare with \gl{herm}).
Hence, according to the "Strip-off" Theorem the matrix
$\tilde{\bf S}^-_{12}$ on the basis of zero-modes of the
operator $b^+_{12}$ contains two (and evidently no more) Jordan cells
of the same size for each eigenvalue.

Next, from the Theorems of Sec.~2 and 6 one concludes that the spectrum
of the matrix $\tilde{\bf S}^-_{12}$ joins the spectra of
$\tilde{\bf S}^\pm_{1}$ and
$\tilde{\bf S}^\mp_{2}$
for the operators $\tilde q^\mp_1$ and  $\tilde q^\pm_2$. Moreover since
the latter ones are stripped off the related matrices  $\tilde{\bf S}^\pm_{i}$
include one Jordan cell only for each eigenvalue.

At last, taking into account that $\mbox{\rm ker}(\tilde q^-_i)\subset
\mbox{\rm ker}(\tilde q^+_j \tilde q^-_i)$ one can derive that for each
eigenvalue of  $\tilde{\bf S}^+_{i}$ a related  Jordan cell with the same
eigenvalue exists in $\tilde{\bf S}^-_{12}$ and its size is no less
than that one of the Jordan cell in  $\tilde{\bf S}^+_{i}$.

Thus the number and sizes of Jordan cells in matrices
$\tilde{\bf S}^\mp_{1}$ and $\tilde{\bf S}^\mp_{2}$ are the same.\\

As a consequence the orders of differential operators  $\tilde q^+_1$ and
$\tilde q^+_2$ are the same and they can be combined to form another intertwining
operator of a lower order, $\sigma^+ = \tilde q^+_1 - \tilde q^+_2$.
If  $\sigma^+ \not= 0$ one can strip off and normalize this operator
$\sigma^+ \rightarrow \tilde\sigma^+$ then apply the Lemma to the pair
of operators  $\tilde\sigma^+ ,\, \tilde q^+_1$
which symmetry operator is
again trivial. As a result we prove these operators to
have the same order in the contradiction  with our initial construction
unless the operator $\sigma^+ =0$ and thereby
$\tilde q^+_1 = \tilde q^+_2$. The latter completes the
proof of the Theorem.

\medskip

As a corollary of the Theorem one finds that for the stripped-off operators
$\tilde q^+_1$ and  $\tilde q^+_2$ of different order the symmetry
operator $\tilde q^+_1 \tilde q^-_2- \tilde q^+_2 \tilde q^-_1 \not= 0$.

Two other consequences of the Theorem concern the structure
of symmetry operators and their uniqueness:\\
a) any symmetric (self-transposed) symmetry operator $B^+ = (B^+)^t,\,
B^+ h^+ = h^+ B^+$ is a
polynomial of the Hamiltonian, the latter is obtained by substituting a pair of
the operators  $B^+$ and $1$ (unity)  instead of  $q^\pm_i,\, i=1,2$ in the
formulation of the Theorem;\\
b) any two antisymmetric symmetry operators $e^+_i = - (e^+_i)^t,\,
e^+_i h^+ = h^+ e^+_i,\, i=1,2$ are dependent, {\it i.e.} being stripped off
coincide, that follows from the Theorem after substituting them  instead
of  $q^\pm_i,\, i=1,2$.

Now one can answer the question about a maximal number of independent
supercharges and their relative oddness. First let us prove that the
number of supercharges cannot exceed two. Assume that there exist
three independent intertwining operators $q^+_i,\, i =1,2,3$.
Then their pairwise symmetry operators  $q^+_i q^-_j - q^+_j q^-_i, \, i\not=j$
 are antisymmetric and non-trivial but dependent in virtue of the
consequence b),  $q^+_i q^-_j - q^+_j q^-_i = \alpha_{ij}(h^+) \tilde e^+$.
When multiplying these relations on  $\alpha_{kl}$ one can assemble
the following
identity,
\ba
&&q^+_1[  \alpha_{13}(h^-) q^-_2 -  \alpha_{12}(h^-) q^-_3] -
[q^+_2 \alpha_{13}(h^-) - q^+_3 \alpha_{12}(h^-)] q^-_1 = 0
=q^+_1 q^-_{23} - q^+_{23}  q^-_1;\no
&&q^+_2 \alpha_{13}(h^-) - q^+_3 \alpha_{12}(h^-) \equiv q^+_{23},
\ea
where the operator  $q^+_{23}$ is non-trivial due to independence of
$q^+_2$ and $q^+_3$.
Evidently two intertwining operators $q^+_1$ and  $q^+_{23}$
satisfy the requirements of the Theorem
and therefore are dependent in contradiction with the initial assumption.
Thus we have proved that the maximal number of independent supercharges is two.

Next let us consider two normalized independent intertwining 
operators   $q^+_i,\, i =1,2$ of orders
$N_1$ and $N_2$ such that  $N_1 > N_2$ and
the sum of their order $N_1 +N_2$ is even. Then evidently the operator
\be
q^+_3 =q^+_1 - (- h^+)^{\frac{N_1 -N_2}{2}} q^+_2 \la{reduc}
\ee
is independent of  $q^+_2$ and has the order   $N_3$ less than  $N_1$.
If the sum of orders  $N_2 +N_3$ is again even one may 
normalize $q^+_3$ and apply the above algorithm to derive a lower order independent SUSY generator until the sum of orders
became odd. Thus one can always construct the basis of two
intertwining operators
containing an even and an odd one.

Finally one can search for a set of minimal intertwining operators $k_N^\pm,
p_{N_1}^\pm$ 
just solving
the system for two independent intertwining operators,
\be
q^\pm_i = \alpha^\pm_i (h^\pm) k_N^\pm + \beta^\pm_i (h^\pm) p_{N_1}^\pm,
\quad i=1,2\ ; \la{decom}
\ee
with coefficients  $\alpha^\pm_i (h^\pm),\, \beta^\pm_i (h^\pm)$ polynomial
in the Hamiltonian.

Indeed,\\
a) among all intertwining operators $q^+$
there exist an unique real operator
of lowest order $p^+$ normalized according to \gl{superch};\\
b)  among all intertwining operators $q^+$ independent of  $p_{N_1}^+$
there exist a real operator
of lowest order $k_N^+$ normalized according to \gl{superch};\\
c) with the help of the algorithm \gl{reduc} one can prove that
one of the operators  $p_{N_1}^+$ and $k_N^+$ is of even order
and another one is of odd order;\\
d) an arbitrary intertwining
operator $q^\pm$ is
always decomposed in the form \gl{decom} in the unique way
which can be proven by a consequent application of the algorithm \gl{reduc}
and taking into account that the one of the
terms  $\sim p_{N_1}^\pm$ and  $\sim k_N^\pm$ is of even order
and another one is of odd order.

Thus the set of  $p_{N_1}^+$ and $k_N^+$ form an optimal basis of
intertwining operators. As  all $q^- =(q^+)^t$ the same results are 
translated
to the set of  $p_{N_1}^-$ and $k_N^-$.
\section{More about symmetry operators}
\hspace*{2ex} In the previous Section we have proven that the antisymmetric 
symmetry
operator (in each component)
 is unique after being stripped off. But the
optimization of supercharge basis may not guarantee the minimal form of
components of the symmetry operator.
The uniqueness of decomposition \gl{decom} allows to formulate a necessary
condition for the symmetry operator $e^\pm$ made of the minimal
operators $k_N^\pm,\, p_{N_1}^\pm$ to be stripped off further on.
Namely if a polynomial $f^\pm (h^\pm)$ can be
factorized out of the symmetry operator $e^\pm$ then the same polynomial
appears as a multiplier in  $\tilde{\cal P}_N (h^\pm),\,
\tilde{\cal P}_{N_1} (h^\pm),\, b^\pm (h^\pm)$. It follows from the relations,
\ba
&& \mp i e^\pm k_N^\pm = \frac12 (k_N^\pm p_{N_1}^\mp -  p_{N_1}^\pm k_N^\mp)k_N^\pm =
b^\pm (h^\pm) k_N^\pm - \tilde{\cal P}_N (h^\pm) p_{N_1}^\pm,\no
&& \mp i e^\pm p_{N_1}^\pm = \frac12 (k_N^\pm p_{N_1}^\mp -  p_{N_1}^\pm k_N^\mp) p_{N_1}^\pm =
 \tilde{\cal P}_{N_1} (h^\pm) k_N^\pm -b^\pm (h^\pm) p_{N_1}^\pm.
\ea

One can give a more detailed description of
each component of the symmetry operator
for a particular class of potentials with the help of
the Lemma.\\

\noindent
\underline{Lemma}\\
{\it Assume that:\\
a) the Hamiltonian $h_a $ commutes with an
antisymmetric real operator $R_a$ of order $2n+1$ which cannot be
stripped off;\quad b) this Hamiltonian has at least one bound state;\quad
c) the wave function $\Psi_0$ characterizes a bound state with the lowest
energy $E_0$.\\
Then there exist a non-singular Hamiltonian $h_b$, a non-singular
real operator $r_a = \partial - (\Psi_0'/\Psi_0)$ and a non-singular
antisymmetric real operator $R_b$ of order  $2n-1$ which cannot be
stripped off such that:
\ba
&& h_a = r_a^t r_a + E_0;\quad
h_b = r_a r^t_a + E_0;\quad  r_a h_a = h_b r_a;\la{2ham}\\
&& R_a = r_a^t R_b r_a;\quad R_b h_b = h_b R_b.\la{2sim}
\ea
}

\medskip

The proof of relations \gl{2ham} is standard for the SUSY QM \ci{abi}.
The partial factorization of $R_a = \hat R_b r_a$ with a non-singular
differential operator $\hat R_b$ of order $2n$ is provided by the equations
$R_a \Psi_0 = 0$ and $r_a \Psi_0 = 0$.
Evidently it is an intertwining operator,
$h_a \hat R_b = \hat R_b h_b$ due to Eqs.~\gl{2ham}. As the Hamiltonian $h_b$
does not have the level $E_0$ the latter relation entails
$\hat R^t_b \Psi_0 = 0$. Hence the factorization takes place
$\hat R^t_b = R^t_b r_a$ with a non-singular differential operator
$R^t_b$ of order $2n-1$. From the intertwining relations it follows that
the operator $R^t_b$ is a symmetry operator. At last its antisymmetry under
transposition can be easily derived from the similar property of  $R_a$.
The operator $R_b$ cannot be stripped off if Eqs.~\gl{2sim} hold and
the operator   $R_a$ has been stripped off already.

\medskip

From the Lemma one can obtain a certain relationship
between the number of bound states of the Hamiltonian and the structure of the symmetry operator. Namely,
suppose that
the Hamiltonian $h_0$ has  $n$ bound states with energies $E_l, \, E_{l+1} > E_l$
and
commutes with a antisymmetric real
operator $R_0$ of order $2n+1$ which cannot be stripped off.  Then
the (normalized) symmetry operator can be factorized,
\be
R_0 = r^t_0\ldots r^t_{n-1} \ \partial\  r_{n-1}\ldots r_0;\quad
r_l \equiv \partial + \chi_l, \la{canon}
\ee
with non-singular real superpotentials $\chi_l$. Respectively the
ladder (dressing chain) relations hold,
\ba
&& h_{l+1} r_l = r_l h_l;\quad  l = 0,\ldots n-1;\no
&&h_l \equiv r_{l-1} r_{l-1}^t + E_{l-1} =
r^t_l r_l +E_l;\quad l = 1,\ldots n-1;\no
&&h_0 = r^t_0 r_0 + E_0;\quad h_n = r_{n-1} r_{n-1}^t + E_{n-1},
\ea
and the hidden symmetry operators arise for each intermediate Hamiltonian,
\ba
&&R_l = r^t_l\ldots r^t_{n-1}\ \partial\ r_{n-1}\ldots r_l;\quad R_n =
\partial;\no
&& R_l h_l = h_l R_l;\quad l = 0,\ldots n.
\ea
Evidently the Hamiltonian $h_n$ describes a free particle and therefore
the Hamiltonian system with a hidden symmetry can be related to
the free-particle system.

In the case b) of Sec.~4 each component of the
symmetry operator can be also 
represented in the canonical factorized form \gl{canon},
\be
e^\pm =i\Biggl[ \partial^3  - (a +\frac32 V_{1,2})\partial -\frac34 V_{1,2}'
\Biggr] = -i
\left(- \partial - \frac{\Psi_{1,2}'}{\Psi_{1,2}}\right)\,\partial\,\left(
\partial - \frac{\Psi_{1,2}'}{\Psi_{1,2}}\right),
\ee 
with the help
of the bound-state wave functions, $\Psi_{1,2} = C \sqrt{V_{1,2} - \beta^2}$.  
In the case c) the potentials \gl{casec} exemplify the Lemma as one of them is
constant.

One can guess that the above relationship
between the Hamiltonian $h_0$ and the symmetry operator $R_0$ is quite general
because
the algorithm of the Lemma helps to transform the system with $n$ bound states
and with a symmetry operator to  a system with $n-1$ bound states and
a symmetry operator of order lower in two units. After one removes
all bound states with this algorithm the remaining Hamiltonian is still
reflectionless and thereby the scattering coefficients are
trivial corresponding to a free-particle system.

\section{Concluding remarks}
\hspace*{2ex}We have established that:\\
a) for supercharges of finite order the Polynomial SUSY can be
always realized when the supercharges are related by transposition;\\
b) in certain cases (for instance, for complex intertwining operators)
several supercharges may commute with the Super-Hamiltonian which may yield
a non-trivial hidden symmetry of such a system;\\
c) the maximal number of independent
supercharges of type $Q$ (or of type $\bar Q$) 
commuting with a given Super-Hamiltonian is two and
in the case of Extended ${\cal N} =2$ superalgebras
there exists an optimal
set of two real supercharges with components of a
 minimal order in derivatives, 
one of which is
an odd-order operator and another one is an even-order operator;\\
d) for Extended ${\cal N} =2$ superalgebras the hidden
symmetry operator ${\cal E}(H)$ defined in \gl{herm}
is unique up to a multiplier polynomial in the Super-Hamiltonian and among
zero-modes of this operator there are all bound
states of the Super-Hamiltonian.

Finally we mention possible extensions of the Theorems and results of this
paper. First of all it seems to be straightforward to apply them to the
Super-Hamiltonians with complex potentials \ci{acdicom} as the SUSY algebra with
transposition symmetry is well defined for such Super-Hamiltonians.
The application to
matrix Super-Hamiltonians \ci{matrix} is less
trivial but certainly interesting as well as a generalization on
multidimensional QM \ci{abi,ain}.
\section*{Acknowledgments}
\hspace*{2ex} One of us (A.A.) is grateful to F. Cannata and J.-P.
Dedonder for useful discussions. This work was supported by the 
Grant RFBR 02-01-00499.

\end{document}